\documentstyle[aps,prl,twocolumn,floats,citesort,epsf]{revtex}

\newcommand{\eqref}[1]{Eq.~(\protect\ref{#1})}
\newcommand{\figref}[1]{Fig.~\protect\ref{#1}}

\begin{document}

\draft

\wideabs{

\title{
The dynamical dimension of defects
in spatiotemporal chaos
}

\author{
David~A. Egolf
}

\address{
Laboratory of Atomic and Solid State Physics, Cornell University,
Ithaca, NY 14853-2501
}

\date{December 6, 1997}

\maketitle


\begin{abstract} 

Using a new time-dependent measure, we demonstrate
for the first time that each defect in a representative
defect-mediated spatiotemporally chaotic system is associated with one
to two degrees of dynamical freedom.  Furthermore, we show that not
all dynamical degrees of freedom are related to the defects;
additional degrees of freedom are due to underlying phase turbulence.
These results yield a deeper understanding of the dynamical
role of defects 
and provide hope that these complicated systems might be reduced to
simpler descriptions.

\end{abstract}

}

\narrowtext

\section{Introduction}
The motions and interactions of topological defects dominate,
at least visually,
the dynamics of a wide variety of nonequilibrium systems
\cite{Cross93}.  This behavior is
observed in systems as diverse as chemical reactions in
shallow layers (e.g., \cite{Ouyang96}), 
thermal convection in horizontal fluid layers (e.g., \cite{Ecke95}), 
aggregation in slime mold colonies (e.g., \cite{Lee96}),
periodically shaken layers of sand (e.g., \cite{Melo95}), 
and electrical activity waves in
heart tissue (e.g., \cite{Maree97}).
Researchers have proposed that 
the macroscopic behavior of these complicated systems
can be understood by considering a simpler system of interacting
defects \cite{Cross93}, instead of the detailed equations describing
the system.

Some nonequilibrium systems containing defects are also
spatiotemporally chaotic, in that the
spatially and temporally disordered dynamics continues
indefinitely.  These systems are often termed ``defect-mediated''
to emphasize the role of the defects.
For such systems one might
intuitively expect that a measure of the average 
complexity~$D$ of the system
(roughly the minimum number of degrees of freedom to describe
the dynamics) is proportional
to the average number of defects $\langle n_d(t) \rangle_t$ in the system.
More generally, we consider the possibility of a contribution unrelated
to the defects, and we account for the time-dependence of the quantities:
\begin{equation}
D(t) = D_b(t) + D_d\, n(t),
\label{fulldeq}
\end{equation}
where $D_d$ is the average contribution per defect.
The background term~$D_b(t)$ is expected to be non-zero if 
the field between the defects does not simply
respond passively to the motion of the defects.

\eqref{fulldeq} is important to understand for several reasons.
First, insight will be gained into what drives the dynamics
of defect-mediated chaotic systems; i.e., whether the
defects alone determine the dynamics.  This result may be an
important step toward understanding not only the building blocks
of extensive chaos \cite{Egolf94}, but also what is necessary to
control the chaos in these systems.  Second,
we may begin to understand what determines the defect statistics
and, in turn, the long-wavelength characteristics, of
these systems.  Third, if \eqref{fulldeq} is found
to be valid, then separate, but coupled, equations for the
defects and the background might allow a simpler description
of the complete system \cite{Elphick91,Aranson91}.  
This description might be further simplified through a 
Langevin equation approach in which the fluctuating background field
is replaced by a noise term \cite{Yakhot81,Zaleski89,Grinstein96}.

We have studied \eqref{fulldeq}
using a representative defect-mediated
spatiotemporally chaotic system, the two-dimensional
complex Ginzburg-Landau equation \cite{Cross93}:
\begin{equation}
\partial_t u(\vec{x}, t) = u + (1 + i c_1) \vec{\nabla}^2 u
                         - (1 - i c_3) |u|^2 u,
\label{2d-cgl}
\end{equation}
where~$u(\vec{x}, t)$ is a complex-valued field of
size~$L \times L$ with periodic boundary conditions, 
and $c_1$ and $c_3$ are real-valued parameters.
\eqref{2d-cgl} is an experimentally relevant amplitude
equation universally valid near the onset of a Hopf
bifurcation from a stationary homogeneous state to an
oscillatory state.  It is closely related to equations
describing chemical reaction-diffusion systems and 
excitable media such as heart tissue.  For a range of values
of the parameters $(c_1, c_3)$, solutions to \eqref{2d-cgl}
exhibit a defect-mediated spatiotemporally chaotic state in which
pairs of spirals of opposite topological charge are continually created
and annihilated \cite{CoulletGilLega89,Chate96}.  

For a wide range of values of $(c_1, c_3)$ in \eqref{2d-cgl},
we demonstrate for the first time that the complexity~$D$ 
of a defect-mediated spatiotemporally chaotic system is {\sl not}
solely due to the defects.  In fact, the degrees of freedom
can be separated into
two sources, suggesting that \eqref{fulldeq} is an appropriate
description of defect-mediated spatiotemporal chaos.
Using a new measure, we show that each defect is associated
with 1--2 degees of dynamical freedom, and
we provide evidence that the remaining degrees
of freedom are due to the underlying phase dynamics.
This result suggests that the chaotic dynamics might indeed be modelled
using separate, but coupled, equations for the defects 
and the background phase-turbulent field.

\section{Measures of Dynamical Behavior}

We integrated \eqref{2d-cgl} using a pseudospectral method with
time-splitting of the operators, carefully 
testing for convergence with respect to spatial and temporal
resolution and with respect to total integration times.
Typically, we used a timestep of $\Delta t = 0.05$ time units,
a system size of $L = 64$ or $L = 32$ with 2 Fourier modes 
(or, equivalently, 2 spatial points)
per unit of length, and total integration times of about $10^5$
time units.  Defects were found by counting the number of
crossings of the zero contours of the real and imaginary parts
of the field $u(\vec{x}, t)$ in \eqref{2d-cgl}.

\begin{figure}
\centerline{\epsfxsize=3.2in \epsfbox{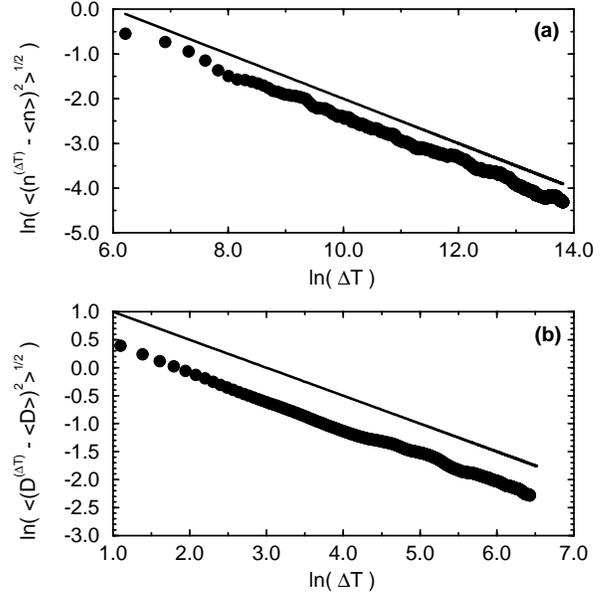}}
\caption{(a) Convergence of the root mean square deviation 
of the finite-time defect density $n^{(\Delta T)}(t)/L^2$
as a function of the time interval~$\Delta T$.
In addition to averages over $t$, results were averaged over 64 different
initial conditions for $(c_1, c_3) = (3.5, 0.90)$ in systems of
size $32 \times 32$ with $\Delta T = 500$. 
(b) Convergence of the root mean square
deviation of the finite-time dimension $D^{(\Delta T)}(t)$
for a system of size $32 \times 32$, $(c_1, c_3) = (3.5, 1.15)$,
and $\Delta T = 32$.
Solid lines are guides to the $\Delta T^{-1/2}$ power law.
}
\label{def-conv}
\end{figure}

The exponential divergence of nearby trajectories in chaotic
systems suggests that two realizations, $u(\vec{x}, t_1)$ and
$u(\vec{x}, t_2)$, of a solution to \eqref{2d-cgl}
can be regarded as statistically independent when the
interval $t_2 - t_1$ is sufficiently large.
Then, for intervals of duration~$\Delta T$
such that $\Delta T$ covers many statistically
independent realizations of
the solution, we expect that the root mean square
deviation of quantities that are each averaged
over intervals of duration~$\Delta T$
will converge as $\Delta T^{-1/2}$.
This conjecture is supported by \figref{def-conv}(a) showing the
convergence of measurements of the average defect density,
$n^{(\Delta T)}(t)$ as a function of the
duration~$\Delta T$ over which each measurement is averaged.
(Each measurement $n^{(\Delta T)}(t)$ is averaged over the
interval $t \leq t' \leq t + \Delta T$.)
The measurements $n^{(\Delta T)}(t)$ converge according
to the form 
\[
\left< \left( n^{(\Delta T)}(t) - 
	\left< n^{(\Delta T)}(t) \right>_t
	\right)^2 \right>_t^{1/2} 
 \propto \Delta T^{-1/2},
\]
where $\langle \rangle_t$ indicates an ensemble average indexed
by the time $t$ for the solution $u(\vec{x}, t)$.

To characterize the dynamics of the chaotic states, we calculated
the time-dependent spectrum of finite-time Lyapunov exponents
($\lambda^{(\Delta T)}_1(t) \geq \lambda^{(\Delta T)}_2(t) \geq 
\cdots$) describing the average rate of divergence 
during time intervals $t \leq t' \leq t + \Delta T$ for
trajectories in phase space that are initially separated by an
infinitesimal amount \cite{Ott93}.
The finite-time exponents were obtained by an
expensive but widely-used algorithm \cite{Parker89}
that involves integrating \eqref{2d-cgl} together with
many copies of the linearization of this equation
about the solution~$u(\vec{x}, t)$.

In previous work on spatiotemporally chaotic systems,
researchers \cite{Manneville85,Egolf94,Egolf95,Ohern96}
have focused on the infinite-time-averaged Lyapunov
exponents (and the closely related Lyapunov dimension) which
average over the fluctuations in the dynamics.
In this work, we utilize the natural fluctuations
within the dynamics to extract additional information about
the system without additional computational expense.
(Because the Lyapunov exponents describe behavior in the
linear regime about the solution, a standard method \cite{Parker89}
for obtaining the infinite-time exponents
is to average the appropriately ordered
finite-time Lyapunov exponents.)
The fluctuations in the complexity of the dynamics is
reflected in changes in the spectrum of finite-time
Lyapunov exponents.  
In analogy to the Kaplan-Yorke conjecture for
(infinite-time) Lyapunov exponents \cite{Cross93}, we 
reduce the spectrum to a single quantity by defining a new
time-dependent measure of the dynamical complexity,
the finite-time dimension $D^{(\Delta T)}(t)$:
\[
D^{(\Delta T)}(t) \equiv k^{(\Delta T)}(t) 
	- \left(\sum_{i=1}^{k^{(\Delta T)}(t)} 
	\lambda^{(\Delta T)}_i(t)\right) 
	\left(\lambda^{(\Delta T)}_{k^{(\Delta T)}(t)+1}(t)\right)^{-1},
\]
where $k^{(\Delta T)}(t)$ is the largest integer such that
$\sum_{i=1}^{k^{(\Delta T)}(t)} \lambda^{(\Delta T)}_i(t) \geq 0$.

\figref{def-conv}(b) suggests that sufficiently separated
values of this new measure $D^{(\Delta T)}(t)$
can be regarded as statistically  independent.
We also note that in the limit $T \rightarrow \infty$,
$D^{(\Delta T)}(t)$
approaches the (infinite-time) Lyapunov dimension~$D_L$.
All results are obtained in
the extensive chaos regime for which $D_L$ 
and $\langle n(t) \rangle_t$ grow
linearly with system size $L^2$.

\section{The dimension per defect}

\begin{figure}
\centerline{\epsfxsize=3.1in \epsfbox{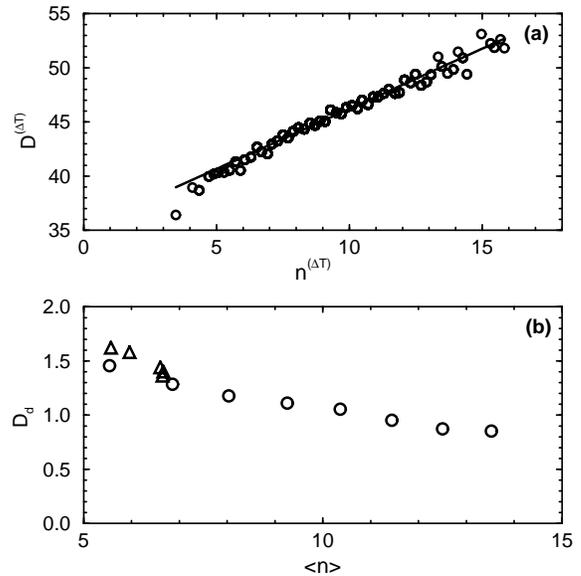}}
\caption{(a) Finite-time dimension $D^{(\Delta T)}(t)$ 
vs. the average number
of defects $n^{\Delta T}(t)$ during the same 
interval~$\Delta T = 32$~time units for 
$(c_1, c_3) = (3.5, 1.15)$ in a system of size $32 \times 32$.
Each point is an average over all measurements
with $n^{\Delta T}(t)$ within
bins of width $0.2$~defects.  The solid line is a linear fit with
each point weighted by the number of measurements in the respective bin.
(b) Values of $D_d$ for various parameters $(c_1, c_3)$.
Circles indicate $c_1 = 3.5$ and
$0.85 \leq c_3 \leq 1.55$; triangles indicate $c_3 = 0.9$
and $2.5 \leq c_1 \leq 5.0$.
}
\label{onedefdim}
\end{figure}

Our new measure, the finite-time dimension $D^{(\Delta T)}(t)$,
allows the comparison of fluctuations in the complexity
to fluctuations in the number of defects (for each set of
system parameters), effectively separating
the degrees of freedom correlated with the defects from the
remaining degrees of freedom.  Using our calculation
of these time-varying quantities, we directly test \eqref{fulldeq}.
For a representative set of the 
system parameters, $(c_1 = 3.5, c_3 = 1.15)$,
\figref{onedefdim}(a)
shows the finite-time dimension $D^{(\Delta T)}(t)$
plotted as a function of the average number of
defects $n^{(\Delta T)}(t)$ found
{\sl during the same time interval} of duration~$\Delta T$.
Each point in \figref{onedefdim}(a) represents an average of the
values of $D^{(\Delta T)}(t)$ for which $n^{(\Delta T)}(t)$
is within a small range of size~0.2 defects.
According to \eqref{fulldeq}, the slope of the fit 
yields $D_d$, the
average number of degrees of freedom associated
with each defect.  The average number of
degrees of freedom not correlated with the defects,
$\langle D_b \rangle_t$ in \eqref{fulldeq}, 
is a significant fraction of the the total dimension.  
(The relatively large contributions of $\langle D_b \rangle_t$ 
to the total dimension
can provide misleading information concerning the dimension per
defect if one simply compares the (infinite-time) Lyapunov dimension
to the number of defects.)
For a wide range of values of finite-time
interval~$\Delta T$ tested (16 time units $\leq \Delta T \leq$
256 time units), the values of the slope and intercept are approximately
constant to within measurement errors, with the range of
values of $D^{(\Delta T)}(t)$ and $n^{(\Delta T)}(t)$ 
decreasing with increasing~$\Delta T$.  The accuracy at larger
values of $\Delta T$ is severely restricted by the decrease in
the number of measurements and the narrowness of
the dimension and defect distributions.

Figures similar to \figref{onedefdim}(a)
are obtained for a wide range of parameters $(c_1, c_3)$.
\figref{onedefdim}(b) shows the values of $D_d$ obtained from
linear fits of the data for each set of system parameters.
We find that $D_d$ decreases slowly as the number of defects increases.
We note that a rough extrapolation of the data in \figref{onedefdim}(b)
toward zero defects suggests an approximate value of
$D_d \approx 2$.

Further work (at great computational expense) will be necessary
to discern the physical origin of the measured values of $D_d$.
The value of $D_d$ may indicate the average fractal dimension of
the trajectories of defects; for example, a value of $D_d = 2$
would be expected if the trajectory of a defect is essentially
random in two spatial dimensions.
Possible explanations for the decrease in $D_d$ at larger defect
densities are correlated behavior between defects and 
constraint of the defect trajectories due to the proximity
of other defects.  Evidence of these possibilities might
be found in the small amount of negative curvature observed
in \figref{onedefdim}(a) and similar plots for other parameter values.
However, much more data will be needed to measure the curvatures
accurately and to assess the impact of
these small curvatures on other quantities reported here.
The small
magnitudes of the curvatures are consistent with
analytical and numerical work showing that defects in \eqref{2d-cgl}
are weakly interacting \cite{Aranson91}.

We are currently investigating these ideas, and we are
studying whether similar values of $D_d$
are found in other defect-mediated spatiotemporally chaotic systems.

\section{The phase contribution}
\begin{figure}
\centerline{\epsfxsize=3.2in \epsfbox{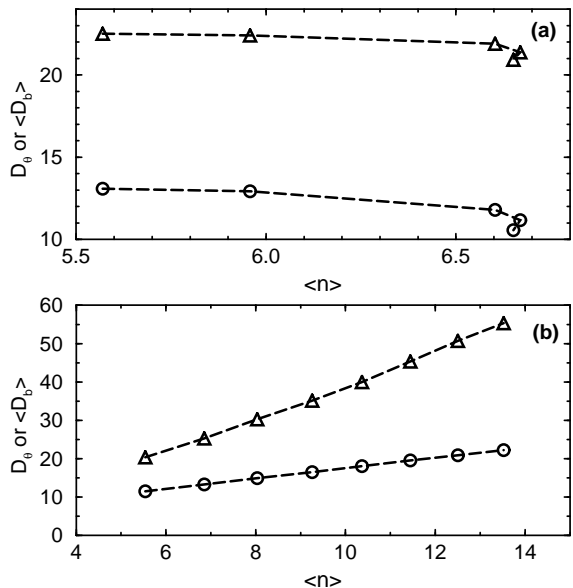}}
\caption{Background dimension~$\langle D_b \rangle_t$ 
(triangles) and phase dimensions~$D_\theta$
(circles) for (a) $c_3 = 0.9$ and $2.5 \leq c_1 \leq 5.0$, and
(b) $c_1 = 3.5$ and $0.85 \leq c_3 \leq 1.55$.
}
\label{backfig}
\end{figure}

The non-defect degrees of freedom~$\langle D_b \rangle_t$ 
vary significantly over
our range of system parameters, as seen in \figref{backfig}.
We speculate that this ``background'' contribution is due to 
an underlying phase turbulence.
To elucidate this further, we obtained an estimate
of $\langle D_b \rangle_t$ using a phase equation 
derived in the limit $\nu = (c_1 c_3 - 1) \rightarrow 0$
\cite{Kuramoto76}.  Upon rescaling, the phase equation can be
written in a parameter-less form known as the
two-dimensional Kuramoto-Sivashinsky equation:
\begin{equation}
\partial_t \theta(\vec{x}, t) =
- \vec{\nabla}^2 \theta - (\vec{\nabla} \theta)^2 
- \vec{\nabla}^2 \vec{\nabla}^2 \theta.
\label{KS-eq}
\end{equation}
The scaling leading to \eqref{KS-eq} introduces a factor
\[
\alpha = \left( \frac{2}{c_1 (c_1 + c_3)} \right) ^ {1/2}
(c_1 c_3 - 1) ^ {1/2},
\]
allowing the computation of a ``phase dimension''~$D_\theta$
of \eqref{2d-cgl} as a function of $(c_1, c_3)$ and the system 
area~$A_{\rm CGL} = L \times L$, given a {\sl single}
measurement of the (infinite-time) dimension $D_{\rm KS}$ of \eqref{KS-eq}
for a system of size $A_{\rm KS}$ \cite{Egolf95}:
\begin{equation}
\frac{D_\theta}{A_{\rm CGL}} = \frac{D_{\rm KS}}{A_{\rm KS}} \alpha^2.
\label{backprediction}
\end{equation}

Remarkably, as seen in \figref{backfig}, the behavior of
the phase dimension~$D_\theta$ as a function of the average
number of defects $\langle n(t) \rangle_t$ is similar to
the behavior of the measured average background
dimension~$\langle D_b \rangle_t$ (although the prediction
of \eqref{backprediction} is smaller than the measured values).
This is particularly evident in the non-trivial behavior on 
the right-hand side of \figref{backfig}(a).
This result is particularly surprising 
since \eqref{KS-eq}
is derived from an expansion in the small parameter
$\nu = (c_1 c_3 - 1) \rightarrow 0$,
while, for the solutions studied here, $\nu = {\cal O}(1)$.
Additionally, since the defects are not bound in pairs of
opposite topological charge, the phase of the complex field
should be globally influenced by the presence of the defects.
The effect of the defects on the phase dimension might not
be as large as one would naively expect since chaotic lengthscales
have been shown to be much shorter than lengthscales corresponding
to the phase \cite{Egolf94,Egolf95,Ohern96}; however, 
this effect might account for the increased difference between
$D_\theta$ and $\langle D_b \rangle_t$ when the system contains more 
defects, as seen in \figref{backfig}(b).
That similarities remain even in the presence of these effects
suggests that the background dimension is closely
related to the phase turbulence.  

Additional insight might also be gained by studying the
solutions to \eqref{KS-eq} and the solutions to \eqref{2d-cgl}
for some small values of $\nu$.  
These solutions are purely phase turbulent 
cellular patterns \cite{Chate96}
with the cells delineated by shocks in the phase
field.  A cellular pattern also can be found in the defect-turbulent
state of \eqref{2d-cgl} with the defects moving predominantly
within the cells.  Further work will be required to test whether
the number of these cells is related to the background dimension.

\section{Conclusions}

Prompted by measurements of the ergodic nature of
spatiotemporally chaotic 
attractors, we have defined a new time-dependent
measure, the finite-time dimension.
By employing this measure we have found a
linear relationship between the degrees of dynamical
freedom and the number of
defects in the system during the same time interval.  These
results show 
that each defect in a defect-mediated spatiotemporal chaotic system
is associated with $D_d \approx $ 1--2 degrees of freedom.
However, a large number of degrees of freedom are not associated
with the defects, and we have provided evidence that these
degrees of freedom are related to the underlying
phase turbulence.  Our results yield a deeper understanding
of defect-mediated chaos and provide hope that these
complicated systems might be reduced to simpler descriptions.


We thank E. Bodenschatz, I. Melnikov, and H. Riecke for valuable
discussions.  This work was supported by the US National Science
Foundation and the Cornell Theory Center.



\begin{thebibliography}{10}

\bibitem{Cross93}
M.~C. Cross and P.~C. Hohenberg, Rev. Mod. Phys. {\bf 65},  851  (1993).

\bibitem{Ouyang96}
Q. Ouyang and J.-M. Flesselles, Nature {\bf 379},  143  (1996).

\bibitem{Ecke95}
R.~E. Ecke, Y. Hu, R. Mainieri, and G. Ahlers, Science {\bf 269},  1704
  (1995).

\bibitem{Lee96}
K.~J. Lee, E.~C. Cox, and R.~E. Goldstein, Phys. Rev. Lett. {\bf 76},  1174
  (1996).

\bibitem{Melo95}
F. Melo, P. Umbanhower, and H.~L. Swinney, Phys. Rev. Lett. {\bf 75},  3838
  (1995).

\bibitem{Maree97}
A.~F.~M. Maree and A.~V. Panfilov, Phys. Rev. Lett. {\bf 78},  1819  (1997).

\bibitem{Egolf94}
D.~A. Egolf and H.~S. Greenside, Nature {\bf 369},  129  (1994).

\bibitem{Elphick91}
C. Elphick and E. Meron, Physica D {\bf 53},  385  (1991).

\bibitem{Aranson91}
I.~S. Aranson, L. Kramer, and A. Weber, Physica D {\bf 53},  376  (1991).

\bibitem{Yakhot81}
V. Yakhot, Phys. Rev. A {\bf 24},  642  (1981).

\bibitem{Zaleski89}
S. Zaleski, Physica D {\bf 34},  427  (1989).

\bibitem{Grinstein96}
G. Grinstein, C. Jayaprakash, and R. Pandit, Physica D {\bf 90},  96  (1996).

\bibitem{CoulletGilLega89}
P. Coullet, L. Gil, and J. Lega, Phys. Rev. Lett. {\bf 62},  1619  (1989).

\bibitem{Chate96}
H. Chat\'{e} and P. Manneville, Physica A {\bf 224},  348  (1996).

\bibitem{Ott93}
E. Ott, {\em Chaos in Dynamical Systems} (Cambridge U. Press, New York, 1993).

\bibitem{Parker89}
T.~S. Parker and L.~O. Chua, {\em Practical Numerical Algorithms for Chaotic
  Systems} (Springer, New York, 1989).

\bibitem{Manneville85}
P. Manneville,  in {\em Liapounov exponents for the {Kuramoto-Sivashinksy}
  model}, Vol.~230 of {\em Lecture Notes in Physics} (Springer-Verlag, ADDRESS,
  1985), pp.\ 319--326.

\bibitem{Egolf95}
D.~A. Egolf and H.~S. Greenside, Phys. Rev. Lett. {\bf 74},  1751  (1995).

\bibitem{Ohern96}
C.~S. O'Hern, D.~A. Egolf, and H.~S. Greenside, Phys. Rev. E {\bf 53},  3374
  (1996).

\bibitem{Kuramoto76}
Y. Kuramoto and T. Tsuzuki, Prog. Theo. Phys. {\bf 65},  356  (1976).

\end{thebibliography}
\end{document}